\documentclass[useAMS,usenatbib]{mn2e}

\usepackage{epsfig}
\usepackage{amssymb}
\usepackage{aas_macros}

\def \apjs {ApJS}
\def \apjl {ApJL}
\def \aap {A\&A}

\begin{document}

\newcommand\phn{\phantom{0}}%
\newcommand{\HB}{H~{$\beta$}}
\newcommand{\Mgg}{Mg~{\sc II}}
\newcommand{\CIV}{C~{\sc iv}}
\def\sarc{$^{\prime\prime}\!\!.$}
\def\arcsec{$^{\prime\prime}\, $}
\def\arcmin{$^{\prime}$}
\def\kms{${\rm km\, s^{-1}}$}
\def\degr{$^{\circ}$}
\def\seco{$^{\rm s}\!\!.$}
\def\ls{\lower 2pt \hbox{$\;\scriptscriptstyle \buildrel<\over\sim\;$}}
\def\gs{\lower 2pt \hbox{$\;\scriptscriptstyle \buildrel>\over\sim\;$}}

\def\mbh{$M_{\rm BH}$}
\def\ergs{$\rm erg\, s^{-1}$}
\def\lam{$\lambda$}
\def\PL{$p(\lambda,z)$}
\def\mstar{$M_{\rm STAR}$}
\def\msun{$M_{\odot}$}
\def\sis{$\sigma$}
\def\vvir{$V_{\rm vir}$}
\def\ms{$M_{\rm STAR}$}
\def\rhoz{$\rho_{\bullet}(z)$}
\def\rhovdf{$\rho_{\rm VDF}(z)$}

\def\gsim{\;\rlap{\lower 2.5pt
 \hbox{$\sim$}}\raise 1.5pt\hbox{$>$}\;}
\def\lsim{\;\rlap{\lower 2.5pt
   \hbox{$\sim$}}\raise 1.5pt\hbox{$<$}\;}


\title[Optical and Radio Quasars]
{The relative growth of Optical and Radio Quasars in SDSS}

\author[F. Shankar et al.]
{Francesco Shankar$^{1}$\thanks{E-mail:$\;$shankar@mpa-garching.mpg.de}, Gregory R. Sivakoff$^{2}$, Marianne Vestergaard$^{3,5}$, and Xinyu Dai$^{4}$\\
$1$ Max-Planck-Instit\"{u}t f\"{u}r Astrophysik,
Karl-Schwarzschild-Str. 1, D-85748, Garching, Germany\\
$2$ Department of Astronomy, University of
Virginia, Charlottesville, VA 22904-4325\\
$3$ Dark Cosmology Centre
Niels Bohr Institute at Copenhagen University
Juliane Maries Vej 30
DK-2100 Copenhagen O, Denmark\\
$4$ Astronomy
Department, University of Michigan, Ann Arbor, MI, 48109\\
$5$ Steward Observatory
933 N Cherry Avenue
Tucson, AZ 85718}


\date{}
\pagerange{\pageref{firstpage}--\pageref{lastpage}} \pubyear{2008}
\maketitle

\label{firstpage}

\begin{abstract}
We cross-correlate the SDSS DR3 quasar sample with FIRST and the Vestergaard et
al. black hole (BH) mass sample to compare the mean accretion histories of optical
and radio quasars. We find significant statistical evidence that radio quasars have a
higher mean Eddington ratio \lam\ at $z > 2$ with respect to optical quasars, while the
situation is clearly reverse at $z < 1$. At $z > 2$ radio quasars happen to be less massive than optical quasars; however, as redshift decreases radio quasars appear
in increasingly more massive BHs with respect to optical quasars. These two trends
imply that radio sources are not a mere random subsample of optical quasars.
No clear correlation between radio activity and BH mass and/or accretion rate is
evident from our data, pointing to other BH properties, possibly the spin, as the driver
of radio activity. We have checked that our main results do not depend on any evident bias.
We perform detailed modelling of reasonable accretion histories for optical and radio quasars, finding that radio
quasars grow by a factor of a few, at the most, since $z\sim 4$. The comparison between the
predicted mass function of active radio quasars and the observed optical luminosity
function of radio quasars, implies a significantly lower probability for lower mass BHs
to be radio loud at all epochs, in agreement with what is observed in the local universe.
\end{abstract}


\section{Introduction}
\label{sec|intro}


\citet{Jiang07} have recently determined that $\sim 10\%$ of
optically selected quasars in the Sloan Digital Sky Survey (SDSS)
are radio-loud, here meaning that they have enough radio power to be
detected in the Faint Images of the Radio Sky at Twenty centimeters
(FIRST; \citealt{Becker95}) survey. This fraction seems to depend on
luminosity and redshift; however, it is still unclear why and how
only a minority of Active Galactic Nuclei (AGNs) show signatures of
powerful radio emission. The simplest scenario usually invoked is
the evolutionary one (e.g., \citealt{Rees84}), where all AGNs
experience a brief radio-loud phase.
Within this interpretation the radio phase of quasars must therefore occur on an
overlapping time-scale much shorter than the optical phase, to explain the small
fraction of radio-loud sources within optical samples.
\citet{Bird08} by matching analytical model predictions
to observed source sizes, have recently
found the radio-jet time-scales to be on the order of $\sim 10^7$ yr,
significantly shorter than the optical time-scale for
quasars, constrained to be $\gtrsim 5\times 10^7$ yr from
demographic arguments
\citep[e.g.,][]{Marconi04,Shankar04,SWM,YuLu08}. However, the
\citeauthor{Bird08}
findings may only probe a single phase of
radio-loudness, while AGNs may undergo several of these events.
Although still limited by the poor knowledge of a comprehensive
census for the radio-loud AGN population, preliminary demographic
studies point towards longer cumulative time-scales for the radio
phase \citep[e.g.,][]{MerloniHeinz08, Shankar08Cav}.

Theoretically, the origin of AGN radio-loudness still constitutes an open
issue. Empirically, still no clear, strong correlation
between radio-loudness and any black hole (BH) property, such as mass, luminosity,
Eddington ratio, or spin has yet been found.
There is tentative evidence suggesting that the formation of a
relativistic jet or a fast wind \citep[e.g.,][]{Blundell07}
sustaining the radio emission is tightly related to the mass of the
central BH \citep[e.g.,][]{Laor00}. \citet{Best05} constructed a
large sample of radio-loud AGNs cross-correlating FIRST, SDSS and
the National Radio Astronomy Observatory Very Large Array Sky Survey
\citep[NVSS;][]{Condon98}, finding that the fraction of radio-loud
AGNs is a strong increasing function of the central BH mass and
galactic stellar mass (\mstar).

Although BH mass might indeed represent an important
aspect regulating radio-loudness, it cannot
be the only key driver. The wide scatter observed between radio and optical
luminosities \citep[e.g.,][]{Ciras03}, and radio power and Eddington
ratios \citep[e.g.,][]{Marchesini04,Sikora07}, for example, suggest that other
parameters such as the mass accretion rate onto the BH and possibly
its spin could also play a major role in determining when a galaxy
becomes radio-loud
\citep[e.g.,][]{Blandford99,Sikora07,GhiselliniTavecchio}.
Theoretical arguments such as those by \citet{BlandfordZ}, also propose
that jets are powered by the extraction
of energy already accumulated in a rotating BH. On the other hand,
the efficiency of the energy extraction from the spinning
BH may not
provide the necessary power for energizing the very luminous
sources. Alternative models \citep[e.g.,][]{Livio99,Cav02radio} have therefore proposed
that a significant fraction of the jet or wind kinetic power must be
directly linked to the rest-mass energy of the currently accreting
matter, thus suggesting some
possible link also between radio power and Eddington ratio \citep[e.g.,][and references therein]{Sikora07,GhiselliniTavecchio}.

Some phenomenological constraints on the nature of jets come from
the observed correlations between radio and kinetic powers, the
latter empirically measured by tracing the integrated $pdV$ work
done by radio AGNs in excavating the cavities observed in the hot
gaseous medium around them
\citep[e.g.,][]{Rawlings91,Willott99,Allen06,Hardcastle07,MerloniHeinz08}.
Knowing the exact kinetic power in AGN jets can provide
important clues on the jet composition, on the origin of the
synchrotron emission, on the relative contributions of
positrons, protons, and Poynting flux to the overall energy budget
\citep[e.g.,][]{BlandfordPayne,Meisenheimer03,Lazarian06},
and on the nature of the jet collimation up to Mpc scales.
\citet{Shankar08Cav} constrained the fraction $g_{\rm k}$ of bolometric luminosity 
turned into kinetic power, by using
an optically selected sample for which
both the optical and the radio luminosity functions were determined.
Given the empirical correlations optical and radio luminosities have with
bolometric and kinetic powers, respectively, they
converted the
optical luminosity function into a radio one.
The match with the radio luminosity
function independently determined for the same sample, yielded $g_{\rm k}\sim 0.10$, with a significant scatter around the
mean, in line with several independent studies \citep[e.g.,][]{Kording08,MerloniHeinz08,CattaneoBest}.
The levels of kinetic power derived from these works is
in agreement with the amount of kinetic feedback required
in theoretical studies of massive galaxies \citep[e.g.,][]{Granato04,Croton06,Cav08}. Also, constraining the kinetic efficiency $g_{\rm k}$ can provide useful constraints
on the duty cycle of radio sources and, in turn, set constraints
on the origin of radio-loudness \citep[see][]{Shankar08Cav,CattaneoBest}.

In addition, understanding the main physical processes that make an AGN
radio-loud is of key importance for assessing the true role
AGNs and supermassive BHs played in the evolution of galaxies. It
has now been proven, in fact, that most, if not all, local galaxies
have a BH at their centre, the mass of which is tightly correlated
with the velocity dispersion \sis$\,$ and other bulk properties of
the host galaxy
\citep[e.g.,][]{Ferrarese00,Gebhardt00,Marconi03,Graham07}.
\citet{LiuJiang} found significant evidence that radio-loud AGNs
follow a different \mbh-\sis\ relation than radio-quiet ones, even
after accounting for different selection effects. A similar offset
has been observed for the \mbh-\mstar\ relation \citep{Kim08}.

Semianalytic models of galaxy formation have grown BHs within
galaxies \citep[e.g.,][]{Granato04,Granato06,Monaco07,Croton06,Marulli08} and
showed that the energetic radiative and kinetic back reactions of
AGNs \citep[e.g.,][and references
therein]{Monaco05,Sazonov05,Churazov06,MerloniHeinz08,Ciotti09}
can solve the overcooling problem in massive systems and
contemporarily settle the local relations between BH mass and galaxy
properties. However, numerical hydro simulations and some
theoretical arguments seem to limit the actual need for AGN
feedback, at least in some regimes
\citep[e.g.,][]{MiraldaKoll,Dekel08,Keres08}.

The nature of AGN feedback is still unclear. Although there are
theoretical \citep[e.g.,][]{Murray95,Granato04,Granato06,Vittorini05,Shankar06,Lapi06,Menci08} and empirical arguments \citep[e.g.,][]{DAIBAL,Ganguly07,Shankar08BAL}
in favour of an AGN feedback driven by winds arising from the
accretion disk around the central BH, other models
\citep[e.g.,][]{Silk05,Saxton05,Pipino09} propose that it is instead
a \emph{jet} that is the actual driver for AGN feedback.
A jet can propagate through an inhomogeneous
interstellar medium, forming an expanding cocoon. The interaction of
the outflow with the surrounding protogalactic gas at first
stimulates star formation on a short time-scale, $10^7$ yr or less,
but will eventually expel much of the gas in a wind.

It is therefore clear that understanding AGN radio-loudness from
first principles, can on one hand reveal interesting
features of BH physics, and on the other hand provide constraints on
models for AGN feedback related to the cosmological co-evolution of
BHs and galaxies. In this paper, we use the quasar sample used in
\citet{Shankar08BAL}, which is the result of the cross correlation
between the SDSS Data Release 3 Quasar catalogue and FIRST. By combining it with the BH mass
measurements and bolometric luminosities
presented in \citet{Veste08}, we were able to compare
accretion properties of large samples of optical and radio quasars.
Although optically selected radio quasars may only be a partial
representation of the overall radio population, they have the
enormous advantage of providing us with the knowledge of fundamental
quantities such as the bolometric luminosity and BH mass. We find
significant differences in the accretion histories of radio and
optical quasars at fixed BH mass and redshift already since $z\sim
4$, supporting a clear distinction between these two populations. We
do not find any clear correlation between radio-loudness and BH mass
and/or accretion rate. These results therefore may support a
scenario in which radio quasars are BHs with environments and/or
intrinsic properties (such as the spin) \emph{different} from the
optical quasars.
In separate papers (Shankar et al. 2009, Sivakoff et al. in prep.) we will
investigate further results when distinguishing among
\citet[][FR]{FR74} sources and Broad Absorption Line quasars.
Our aim in these papers is to constrain the differences in accretion
histories for different families of AGNs, thus providing useful
empirical constraints for theoretical models.

This paper is organized as follows. In \S~\ref{sec|data} we describe
the datasets used for the cross-correlations and BH mass estimates.
In \S~\ref{sec|results} we provide our results on the Eddington
ratio, BH mass and redshift distributions of optical and radio
quasars. In \S~\ref{sec|discu} we discuss our findings, in reference
to previous works and give our conclusions in \S~\ref{sec|conclu}.
Throughout this paper we use the cosmological parameters
$\Omega_m=0.30$, $\Omega_\Lambda=0.70$, and $H_0=70\, {\rm
km\, s^{-1}\, Mpc^{-1}}$.
%

\begin{figure*}
\centering
\includegraphics[width=0.9\textwidth]{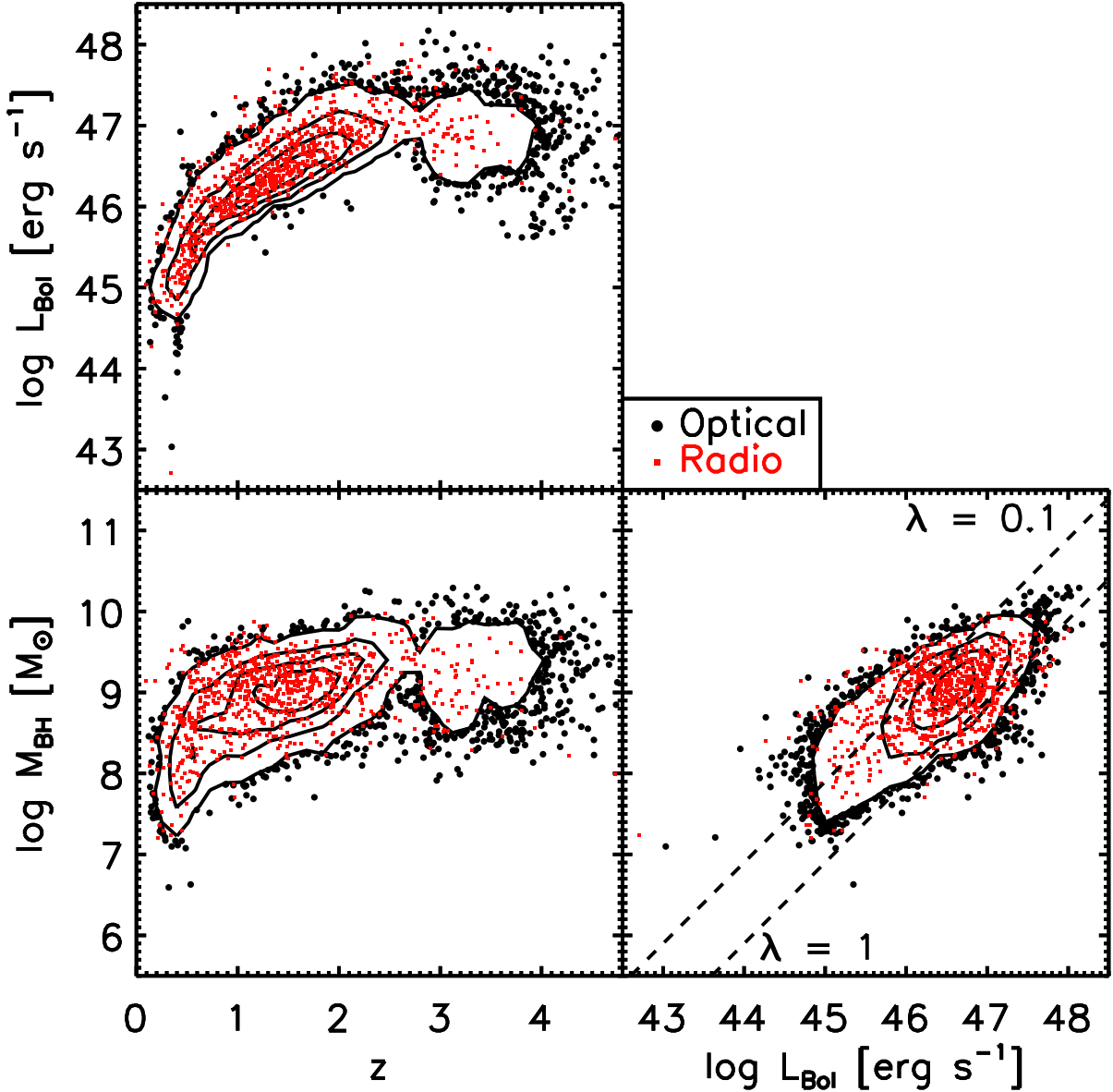}
\caption{Overall distributions of bolometric luminosities,
black hole masses and redshifts for the samples of SDSS quasars with and without
FIRST counterparts, radio and optical quasars, respectively, used in this paper.
The \emph{solid} contours delineate the regions containing 25\%, 50\%, 75\% and
95\% of the optical sample. The \emph{black} dots individually indicate the
remaining 5\% of optical quasars. All radio quasars are shown by the \emph{red}
squares. The optical and radio quasars cover similar areas with no clear
systematic offsets between the distributions. The dashed lines in the lower
right panel show the locus of points in the \mbh-$L$ plane with \lam$=0.1$ and
\lam$=1$, as labelled ($\lambda=L/L_{\rm Edd}$).} \label{fig|initialSummary}
\end{figure*}

\section{DATA}
\label{sec|data}

We adopt the SDSS DR3 quasar catalogue by \citet{Schneider05} as the basis for our
analysis. The data for this sample were taken in five broad optical
bands ($ugriz$) over about 10,000 deg$^2$ of the high Galactic
latitude sky. The majority of quasars were selected for
spectroscopic followup by SDSS based on their optical photometry. In
particular, most quasar candidates were selected by their location
in the low-redshift ($z \lesssim 3$) $ugri$ colour cube with its
$i$-magnitude limit of 19.1. A second higher redshift $griz$ colour
cube was also used with a fainter $i$-magnitude limit of 20.2.

The DR3 quasar catalogue by \citet{Schneider05} also provides the radio flux
density for those sources which have a counterpart within 2\arcsec in the FIRST
catalogue \citep{Becker95}. According to \citet{Schneider05}, while only a small
minority of FIRST-SDSS matches are chance superpositions, a significant fraction
of the DR3 sources are extended radio sources. This can lead to slightly larger
offsets between SDSS and FIRST positions, as well as multiple radio
components. Furthermore, radio lobes may be more strongly offset from the
central optical source. As discussed in \citet{Shankar08BAL}, we built a full
FIRST-SDSS cross-correlation catalogue, containing all the detected radio
components within 30\arcsec of a optical quasar. From this catalogue we define a
radio quasar as any SDSS quasar with either a single FIRST component within
5\arcsec (FRI) or multiple FIRST components within 30\arcsec (FRII). While
here we are mainly interested in identifying all the possible
radio matches within the optical sample, in a
separate paper \citep{Shankar09a} we will focus on the differences in the
intrinsic properties of compact and extended radio sources.
The remaining SDSS quasars that
overlap with FIRST are defined as an optical quasar. In this paper, the radio
sample is restricted to radio quasars whose sum of the integrated flux density
in FIRST $f_{\rm int}$ is above 3 mJy.

\begin{figure*}
\centering
\includegraphics[width=17.5cm]{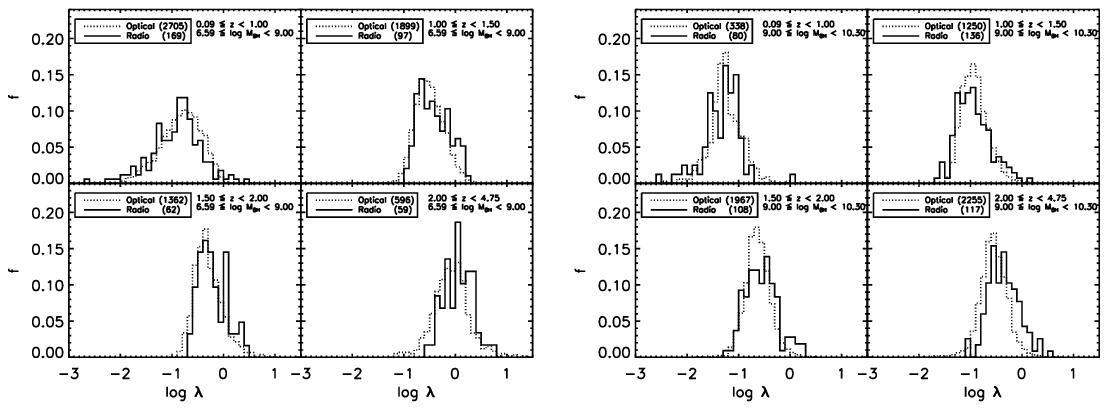}
\caption{\emph{Left panel}: normalized distribution of Eddington
ratios for active black holes with mass in the range
$6.59\lesssim \log M_{\rm BH}/M_{\odot}\lesssim 9.0$ and in
different redshift bins, as labelled.
\emph{Right panel}: normalized distributions of
Eddington ratios for active black holes with mass in
the range $9.00\lesssim \log M_{\rm BH}/M_{\odot}\lesssim 10.30$ and
in different redshift bins, as labelled.
\emph{Solid }lines refer to radio quasars alone, while
\emph{dotted} lines refer to optical quasars only.
Irrespective of the black
hole mass interval considered, the Eddington ratio distributions of
radio quasars are skewed towards higher values of \lam\ at higher
redshifts. In parenthesis we list the number of quasars in each sample.}
\label{fig|LambdaDistributionsLowHighMass}
\end{figure*}

We remind the reader here that FIRST efficiently identifies radio matches to
optically-selected quasars. By cross-correlating SDSS with the large
radio NRAO VLA Sky Survey \citet{Condon98}, \citet{Jiang07}
found in fact that only $\sim$ 6\% of the matched quasars were not
detected by FIRST. 

In this paper we cross-correlate our sample with the one worked by
\citet{Veste08}. The latter estimated the mass function of active
BHs using the quasar sample by \citet{Richards06} with a
well-understood selection function. The reader is referred to
\citet{Richards06} for details on this sample and its selection.
To estimate the mass of the central BH in each quasar,
\citet{Veste08} measured the widths of each of the \HB, \Mgg, and
\CIV\ emission lines, and the monochromatic nuclear continuum
luminosity near these emission lines, with each spectrum corrected
for Galactic reddening and extinction.
When a particular quasar had two emission lines for which it was possible
to determine a BH mass, the final mass estimate was taken to
be the variance weighted average of the individual
emission line based mass estimates. The continuum components were
modelled using a nuclear power-law continuum, an optical-UV iron
line spectrum, a Balmer continuum, and a host galaxy spectrum.
The monochromatic nuclear continuum luminosities near the emission lines were used
to calculate the bolometric luminosity for the quasar.
The
continuum components were subtracted and the emission lines were
then modelled with multiple Gaussian functions so to obtain smooth
representations of the data. All \Mgg\ and \CIV\ profiles with
strong absorption, as identified by visual inspection of the quasar
spectra in the \citet{Trump06} catalogue and of the quasars with redshifts between 1.4, when \CIV\ enters the observing window, and 1.7, were discarded by
\citet{Veste08} from further analysis. Of the 15,180 quasars on
which the DR3 quasar luminosity function is based, BH mass estimates
were possible for 14,434 quasars (95\%).

The bolometric luminosities are based on the fitted nuclear power-law continuum
level extrapolated to $4400$\AA, $L_{\lambda}$, obtained from the spectral decomposition.
The $\lambda L_{\lambda}$ ($4400$\AA) values are scaled by a bolometric correction factor
of $9.20(\pm 0.24)$, determined from the database of \citet{RichardsSED}. We checked that results
do not change when adopting different wavelengths at which luminosities were estimated.

\section{RESULTS}
\label{sec|results}

In Figure~\ref{fig|initialSummary} we show the sample of
optical and radio quasars in slices of the redshift-bolometric luminosity-BH mass plane.
The contours levels delineate the regions containing 25\%, 50\%, 75\% and 95\%
of the optical quasars; individual solid black dots represent the
remaining 5\%. The red squares individually show the radio quasars.
It is clear from this Figure that the radio
and optical samples are well mixed at all luminosities and
redshifts, and there is no apparent selective segregation between
the two types of AGNs. The lower right panel of
Figure~\ref{fig|initialSummary} shows that the bulk of the optical
and radio quasars are strong accretors with Eddington ratios\footnote{We define as Eddington ratio the quantity $\lambda=L/L_{\rm Edd}$, with $L$ the bolometric luminosity and $L_{\rm Edd}=1.26\times 10^{38}\, (M_{\rm BH}/M_{\odot})\, $\ergs.} within
$0.1<\lambda<1$, in line with several independent studies
\citep[e.g.,][]{Veste04,Kollmeier06,Netzer07,Shen08bias}. The
Super-Eddington accretors are a minority \citep[e.g.,][]{MD04},
while there is a non-negligible fraction of sources radiating at
significantly sub-Eddington regimes. To further develop the
comparison between optical and radio sources, we
present a more detailed study in the following section, dividing the optical and
radio samples in bins of redshift, BH mass, and Eddington ratios.


\subsection{EDDINGTON RATIOS}
\label{subsec|comparingLambdas}

\begin{figure*}
\includegraphics[width=17.5truecm]{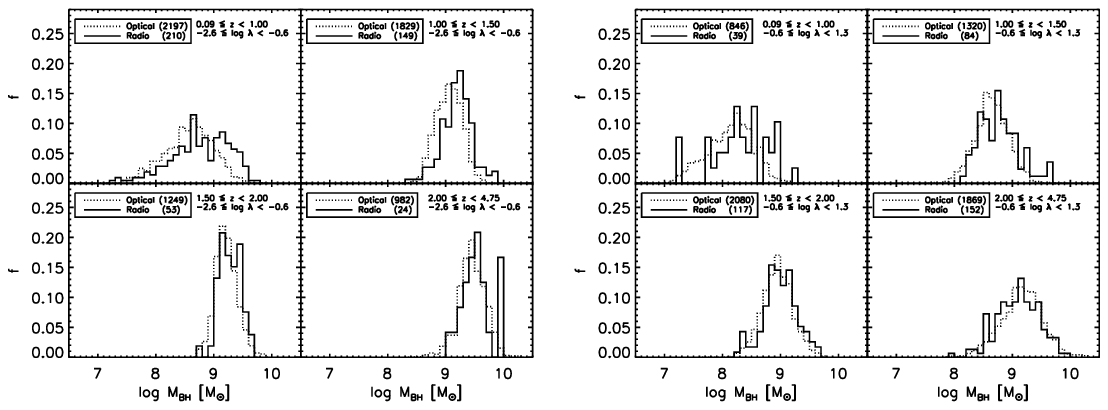}
\caption{\emph{Left
panel}: normalized distributions of black hole mass for active black
holes with Eddington ratio in the range $-2.6\le \log \lambda <
-0.6$. \emph{Right panel}: normalized distributions of black hole mass
for active black holes with Eddington ratio in the range $0.6\le \log
\lambda < 1.3$. In both panels \emph{dotted} lines refer to optical
quasars while \emph{solid} lines to radio ones. While for sources
accreting at low Eddington ratios radio quasars always have higher
black hole masses, on average, there is no clear distinction in the
black hole mass distributions for radio and optical quasars accreting at
higher Eddington ratios. In parenthesis we list the number of quasars in each sample.} \label{fig|MbhDistributionsLowHighLambda}
\end{figure*}

Figure~\ref{fig|LambdaDistributionsLowHighMass} shows normalized
distributions\footnote{Throughout the paper we indicate normalized distributions with the letter \emph{f} in the Figures.} of Eddington ratios for active BHs in different redshift
bins, as labelled. The left panel considers only BHs with mass within
the range $6.59\lesssim \log M_{\rm BH}/M_{\odot}\lesssim 9.0$,
while the right panel considers only the subsamples of more massive
BHs with mass $\log M_{\rm BH}/M_{\odot}>9$. In both panels, the
solid and dotted lines refer to radio and optical quasars,
respectively. We find that the distributions of radio quasars are
clearly skewed towards higher values of \lam\ at redshifts $z>2$.
The distributions get closer at intermediate redshifts $1<z<2$,
while radio quasars shift towards lower Eddington ratios at lower
redshifts. A similar, and even more marked behaviour, is present in
the \lam-distributions of the more massive BHs, plotted in the right
panel of Figure~\ref{fig|LambdaDistributionsLowHighMass}. Radio
quasars accrete at significantly higher Eddington ratios at $z>2$
and then later in time move towards lower and lower \lam s faster
than optical quasars.
We have checked that these results do not depend on the exact choice
of the redshift bins in which we divide the samples, as long as a
significant number of sources for both samples is present in each
bin.

\subsection{BLACK HOLE MASSES}
\label{subsec|comparingBHmasses}

Figure~\ref{fig|MbhDistributionsLowHighLambda} shows the normalized
distributions of BH mass for optical and radio quasars (dotted and
solid lines, respectively) in the same four redshift bins considered
in Figure~\ref{fig|LambdaDistributionsLowHighMass}. The left panel
shows the distribution of only the ``fading'' quasars, i.e., those
shining at low Eddington ratios, in the range $-2.6\le \log \lambda
< -0.6$. We find that, at all times, the radio quasars have a BH
mass distribution peaked at higher masses than optical quasars, on
average. The right panel shows instead that the former trend is not
apparent in highly accreting quasars with $-0.6\le \log
\lambda < 1.3$. Radio quasars exhibit a slight tendency
to have \emph{lower} BH masses
than optical sources at $z>2$, comparable BH masses at intermediate
redshifts $1.5<z<2$, and higher BH masses at lower redshifts.
This shows that
radio quasars in general are not a random subsample
of optical quasars, and have a specific, different
cosmological accretion history than optical quasars.


\subsection{COMPARING THE MEAN VALUES OF THE DISTRIBUTIONS}
\label{subsec|summaryplots}

\begin{figure*}
\includegraphics[width=17.5truecm]{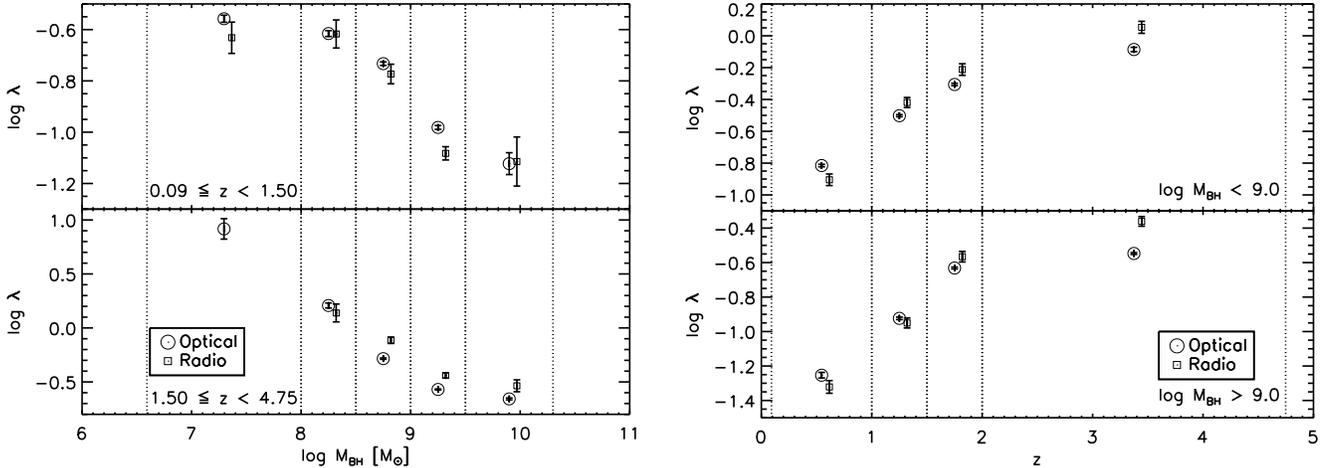}
\caption{ \emph{Left
panel}: mean Eddington ratio as a function of black hole mass for
sources at $z<1.5$ (\emph{upper} plot) and at $z\ge 1.5$
(\emph{lower} plot). \emph{Right panel}: mean Eddington ratio as a
function of redshift for black holes with mass $\log M_{\rm
BH}/M_{\odot}<9.0$ (\emph{upper} panel) and $\log M_{\rm
BH}/M_{\odot}\ge9.0$ (\emph{lower} panel).
In both panels radio and optical quasars are shown with \emph{squares} and
\emph{circles}, respectively. With respect to optical quasars, massive radio
quasars tend to have higher Eddington ratios at higher redshifts and lower or
comparable Eddington ratios at lower redshifts. Also, radio quasars tend to have
higher Eddington ratios at $z\gtrsim 1.5$, and lower Eddington ratios at lower
redshifts. Here and in the following Figures, the vertical \emph{dotted} lines mark the redshift intervals in which the sample was divided, and the results do not depend on the exact choice of such intervals.}
\label{fig|SummaryLambdavsBHmassLambdavsRedshift}
\end{figure*}

In the previous sections we showed that interesting differences
arise when comparing the Eddington ratio and BH mass distributions
for optical and radio sources. The
distributions we find in each subsample considered are always broad,
due to a combination of intrinsic scatter and measurement errors
\citep[e.g.,][]{Shen08bias}. It is therefore worth comparing only the
mean values of the distributions\footnote{We stress that although mean and median quantities do not often coincide in our sample,
our results on the systematic differences between the properties characterizing optical and radio sources are robust against using either of the two.}.
Given the large statistics in our
sample, the mean values are well defined and the errors on the means are
small enough to provide significant constraints. In
this section we therefore summarize our results by comparing the
mean values of the distributions of optical and radio quasars. We
plot these comparisons in four bins of redshift and BH mass chosen
in a way to yield a similar number of sources separately for optical
and radio quasars, as was done for the previous Figures. However,
the results do not depend on how we decide to bin the data. For
example, choosing narrower bins actually enhances the differences
between optical and radio quasars.

With respect to optical quasars, radio quasars tend to
have higher Eddington ratios at redshifts $z>1.5$ while they accrete
at lower, or at most comparable, rates at lower redshifts. The right
panel of Figure~\ref{fig|SummaryLambdavsBHmassLambdavsRedshift}
shows that the mean \lam\ of radio quasars (squares) is
significantly higher than the mean \lam\ of optical quasars at
$z>1.5$, for BHs with mass below (upper plot) and above
(lower plot) $\log M_{\rm BH}/M_{\odot}=9.0$. At lower redshifts all
quasars progressively decrease their mean Eddington ratio, but the
mean \lam\ associated with radio quasars decreases faster, and
eventually becomes lower, than the one associated with optical
quasars. Note that high mass radio quasars with $\log M_{\rm
BH}/M_{\odot}>9$ ``cross'' the optical mean \lam\ around redshift
$z\sim 1.5$, while lower mass radio quasars cross the optical
boundary at later times, around $z\sim 1$.

\begin{figure*}
\centering
\includegraphics[width=17.5truecm]{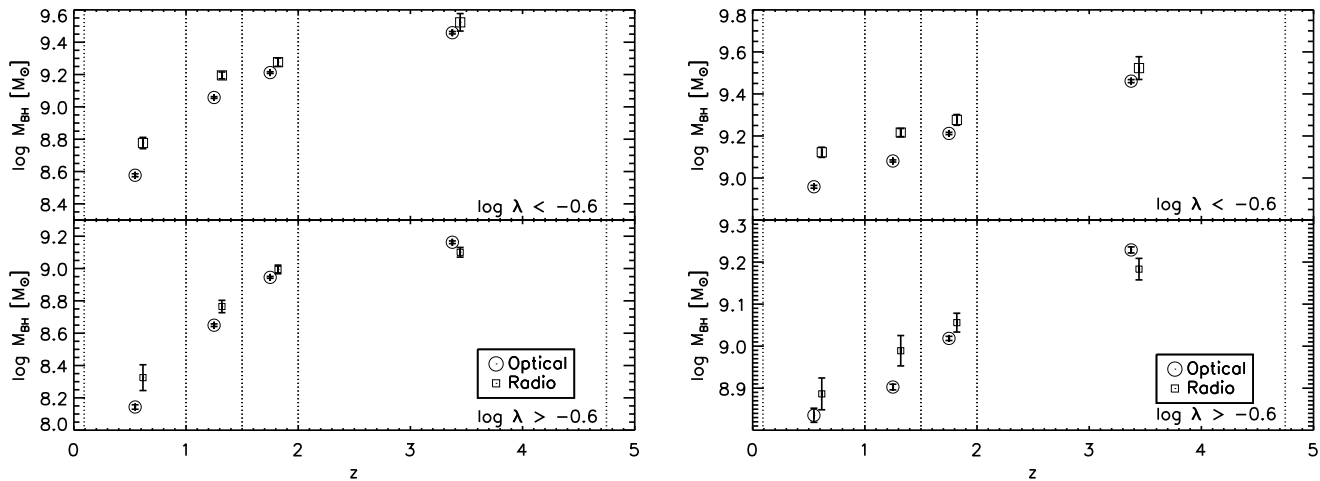}
\caption{\emph{Left panel}: mean black hole mass as a function of
redshift for sources accreting with an Eddington ratio $\log \lambda
<-0.6$ (\emph{upper} plot) and $\log \lambda
>-0.6$ (\emph{lower} plot). \emph{Right panel}: same pattern as
left panel considering only the subsample of sources with black hole
mass higher than $10^{8.7}\, M_{\odot}$, for which no strong
luminosity bias should be present. In both panels optical and radio
sources are shown with \emph{circles} and \emph{squares},
respectively, as labelled. Radio sources with high \lam\ have lower
masses with respect to optical ones at $z>1.5$, but have a tendency for higher
BH masses at lower redshifts.}
\label{fig|SummaryMbhvsRedshiftHighLowLambda}
\end{figure*}

The subsample of sources
with BH masses above $\log M_{\rm BH}/M_{\odot}=9$ is particularly
meaningful. In fact, this subsample suffers from
flux-limited effects much less than the total sample (cf. Figure~\ref{fig|initialSummary}), thus further supporting
the evidence that the differences between optical and radio quasars
are not induced by SDSS selection effects. To be even more conservative, when selecting the sources
with $\log M_{\rm BH}/M_{\odot}>9$, \lam$>-0.6$, and $1<z<2$, to ensure full detectability above the SDSS flux limits, we find that optical and radio sources still differ in their mean \lam\ at the $3.3\sigma$ significance level.

Taken at face value, the data seem to also support a decreasing \lam\ with decreasing redshift,
for both optical and radio sources, in line with previous studies \citep[e.g.,][]{MD04,Veste04,NetzerT07}. When restricting the analysis to the subsample of massive BHs with $\log M_{\rm BH}/M_{\odot}>9$, we still find significant evidence for a decrease in \lam\ ($2.6\sigma$), although the amplitude of the drop is smaller. A decreasing \lam\ with decreasing $z$ is not surprising, given that locally the median Eddington ratio
of all BHs is only a few percent \citep{Kauffmann09}, i.e., at least an order of magnitude lower than what is observed for luminous quasars at high redshifts \citep[e.g.,][]{Veste04,Kollmeier06,Shen08bias}.

The left
panel of Figure~\ref{fig|SummaryLambdavsBHmassLambdavsRedshift} shows instead  significant evidence for a decrease in the mean Eddington
ratio with increasing BH mass for all quasars regardless of radio properties. This trend seems to be
independent of redshift, although at least part of the observed drop might depend on selection effects (see \citealt{Shen08bias}).
In physical terms, this behaviour would naturally arise
if more massive BHs tend to accrete most of their mass at early
times and then undergo a long, ``post-peak'' descending phase
characterized by lower and lower Eddington ratios \citep[e.g.,][and
references
therein]{Granato04,Fontanot06,HopkinsHernquist08a,YuLu08,Ciotti09}.
Many groups claimed a decreasing Eddington ratio with increasing BH mass, using both
brighter and fainter samples than ours \citep[e.g.,][]{MD04,NetzerT07,Kollmeier06,Shen08bias,Dietrich09}. In particular, \citet{Hickox09} recently presented the Eddington ratio distributions of a sample of 585 AGNs at $0.25 < z < 0.8$, finding that radio AGNs, on average the most massive BHs in their sample, have a median \lam\ much lower than AGNs identified in other bands, characterized by lower mass BHs.

The left panel of Figure~\ref{fig|SummaryMbhvsRedshiftHighLowLambda}
shows instead the mean BH mass in the optical and radio samples
as a function of redshift for low and high accretors (upper and
lower plots, respectively). While radio-loud, low accretors always
have higher BH masses with respect to optical quasars, in the lower plot we
see a tendency for radio quasars
to have lower BH masses at $z>2$ and a steady increase to higher
masses at lower redshifts. Therefore, irrespective of how fast
quasars are accreting, at late times radio quasars seem to always be
associated with more massive systems, with the mass difference gradually decreasing with increasing redshifts.
A similar result was also found
by \citet{MetMaglio} using a homogeneous sample of $\sim 300$ radio-loud quasars drawn from the FIRST and 2dF
QSO surveys in the range $0.3 < z < 3$.
To check that these trends are not affected by
flux-limit issues, in the right panel of
Figure~\ref{fig|SummaryMbhvsRedshiftHighLowLambda} we show the same
plots for only the subsample of sources with BH mass above $\log
M_{\rm BH}/M_{\odot}=9$, which fully confirms the trends derived for
the full sample.

To evaluate the significance of our results we performed a
Student T-test (allowing for unequal variances) on the difference between the means of the
distributions in each bin of redshift considered.
The means and standard deviations of the distributions are computed from the biweight mean $\mu$
and biweight standard deviation $\sigma$ \citep{Hoaglin83}. Errors on this mean were
estimated by reducing $\sigma$ by $\sqrt{N}$, where $N$ is the number of quasars in the distribution.
Table 1 summarizes the differences in the mean \lam-distributions
of optical and radio sources, for each bin of redshift and BH mass considered
so far.
In the last column
we report the probability that optical and radio quasars have the same mean,
as determined by the Student's T statistic $P_T$.
It is evident that a clear pattern arises when comparing the \lam-distributions
of the two quasar populations. Irrespective
of the BH mass bin considered, the mean Eddington ratio differs significantly at $z\gtrsim 2$ at the $\sim 3\sigma$ level (i.e., $P_T<2.7\times 10^{-3}$), getting more similar
at moderate redshifts $1.5\lesssim z \lesssim 2$, and differentiating
again at lower redshift at a slightly lower, but still significant, 2$\sigma$ level (i.e., $P_T<4.6\times 10^{-2}$). Table 2 shows that the difference in the median value
of the \mbh-distributions become, on average, significantly more
different when moving from higher to lower
redshifts, and this trend characterizes both high and, to a somewhat lower degree, low-\lam\ accreting BHs.

\subsection{MEAN ACCRETION HISTORIES}
\label{subsec|accretionhistories}

The summary plots and tables discussed in \S~\ref{subsec|summaryplots}, show that significant
differences are present in the \mbh- and \lam-distributions
of optical and radio quasars. In this section, we go a step further and work out their relative expected accretion histories.
To probe the average evolution of the radio and optical quasars of a
given BH mass \mbh, we compute the BH mass function at any time via a continuity equation
\citep[e.g.,][]{Cav71,SB,YT02,Marconi04,YuLu04,Hop07,Shankar09b,SWM,ShankarReview}
\begin{equation}
\frac{\partial n}{\partial t}(M_{\rm BH},t)=-\frac{\partial (\langle
\dot{M}_{\rm BH}\rangle n(M_{\rm BH},t))}{\partial M_{\rm BH}}\,,
    \label{eq|conteq}
\end{equation}
where $\langle \dot{M}_{\rm BH}\rangle=S(M_{\rm
BH},z,\lambda)\langle \lambda\rangle M_{\rm BH}/t_s$ is the mean
accretion rate (averaged over the active and inactive populations, with $t_s=4\times 10^7(\epsilon/0.1)$ yr, with the radiative efficiency $\epsilon=0.1$)
of the optical BHs of mass \mbh\ at time $t$.
Equation~\ref({eq|conteq}) states that the average growth rate of all
BHs is proportional to the function $S(M_{\rm BH},z,\lambda)$, i.e.,
the fraction of BHs of mass $M_{\rm BH}$ active at redshift $z$ and
accreting at the Eddington rate \lam. Equation~(\ref{eq|conteq}) states that every BH, on average, constantly grows at the mean accretion
rate $\langle \dot{M}_{\rm BH}\rangle$ (see \citealt{SteedWeinberg,SWM} for further details). Note that we neglect
any source term in equation~(\ref{eq|conteq}), which may take into
account the (uncertain) BH creation and merger rates. The latter
is a reasonable assumption
given that the overall local BH mass function can be easily accounted for
assuming that most BHs grow through radiatively efficient accretion \citep[see][]{SWM}.

Here we further assume, for simplicity, that the function $S(M_{\rm
BH},z,\lambda)$ can be further separated into
\begin{equation}
S(M_{\rm BH},z,\lambda)=p(\lambda,z)U(M_{\rm BH},z)\, ,
    \label{eq|separatevariable}
\end{equation}
which imposes that all active BHs of mass \mbh\ at redshift $z$, share the same
mean Eddington ratio distribution $p(\lambda,z)$, with $U(M_{\rm
BH},z)$ the duty cycle, i.e., the \emph{total} fraction of active
BHs at redshift $z$ and mass \mbh\ in the BH mass function $n(M_{\rm BH},z)$. We will further discuss
the validity of this assumption. In models with a single value of
\lam, the duty cycle is simply the ratio of the luminosity and mass
functions,
\begin{equation}
U(M_{\rm BH},z)=\frac{\Phi(L,z)}{\Phi_{\rm BH}(M_{\rm BH},z)}\, ,
\,\,\,\,\,\,\,\,\,\,\,\,\,\,\,\, L=\lambda l M_{\rm BH}\, ,
    \label{eq|P0general}
\end{equation}
with $l=1.26\times 10^{38}\, \rm{erg\, s^{-1}\, M_{\odot}^{-1}}$. A physically
consistent model must have $U(M_{\rm BH},z)\le 1$ for all \mbh\ at
all times.

\begin{figure}
\includegraphics[width=7truecm]{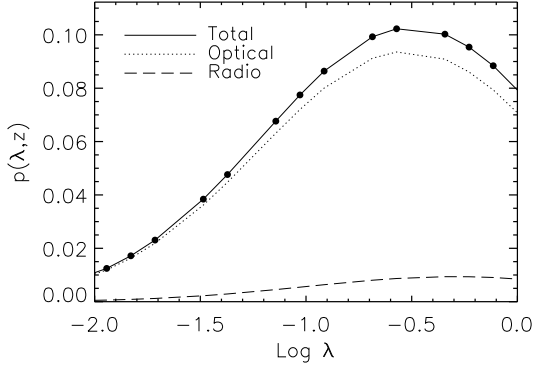}
\centering 
\caption{Eddington ratio distribution adopted in our modelling. The
\emph{solid} line shows the total Gaussian \lam-distribution adopted
for the full population of black holes. The \emph{solid points} mark
the actual discrete values of \lam\ used in the computation. The
\emph{dotted} and \emph{long-dashed} lines represent the separate
distributions for optical and radio sources, respectively. We allow
the peaks of the both Gaussian distributions to decrease in time
following the different evolution for optical and radio sources
shown in the right panel of
Figure~\ref{fig|SummaryLambdavsBHmassLambdavsRedshift}.}
\label{fig|PLambda}
\end{figure}

To model the mean accretion rate we assume that optical and radio
sources have a similar Gaussian shaped Eddington ratio
distribution \PL, but with different means which evolve differently
with time, as given by our results in the right panel of
Figure~\ref{fig|SummaryLambdavsBHmassLambdavsRedshift}.
Unless otherwise noted, we assume that the standard deviation of the Gaussian $\Sigma$ in
\lam\ is 0.7. Although, this value for $\Sigma$ is slightly larger than what we actually observe, it accounts for some of the flux-limited biases discussed by \citet{Shen08bias}. We also note that the exact choice for $\Sigma$ does not alter our results.
We solve Equation~(\ref{eq|conteq}) using the numerical code
discussed in \citet{SWM} and \citet{ShankarReview}, and we refer to those papers for full
details. We briefly point out here that the code computes
the total duty cycle $U(M_{\rm BH},z)$ at any redshift $z$, given the BH
mass function at redshift $z+dz$ and an input \PL\ distribution. The
code allows for any input \PL\ distribution, as long as it is
expressed in discrete form. Figure~\ref{fig|PLambda} shows, for example, the Eddington ratio distribution
adopted in our modelling at $z=3$. The solid line shows the total
\lam-distribution adopted for the full BH population, while the solid circles mark the actual discrete values of \lam\
used in the computation. The dotted and long-dashed lines represent
the separate \PL\ Gaussian distributions for optical and radio sources,
respectively. We allow the median \lam\ values peaks in the \PL\ Gaussian distributions of optical and radio quasars, to decrease with decreasing redshift following the results
in Figure~\ref{fig|SummaryLambdavsBHmassLambdavsRedshift}. The mean accretion rate then has two contributions represented by two terms
\begin{eqnarray}
\langle \dot{M}_{\rm BH}\rangle \propto \left[\int \lambda p_{\rm opt}(\lambda,z) d\lambda+\int \lambda p_{\rm radio}(\lambda,z) d\lambda\right]\times\nonumber\\
M_{\rm BH}\, U(M_{\rm BH},z)\, ,
\end{eqnarray}
the first one $p_{\rm opt}(\lambda,z)$ and the second one, $p_{\rm radio}(\lambda,z)$,
represented by the dotted and long-dashed lines in Figure~\ref{fig|PLambda}. We always assume
$p_{\rm radio}=0.1\times p_{\rm opt}$, to satisfy the empirical constraint that
radio sources are on average 10\% of the optical population \citep[e.g.,][and references therein]{Jiang07}. Note that we are here describing
the radio population as a whole. It may well be true
that compact and extended sources evolve differently
along cosmic time but, as stated above, we leave
this more subtle subdivision for a separate study.

\begin{figure*}
\centering
\includegraphics[width=17.5truecm]{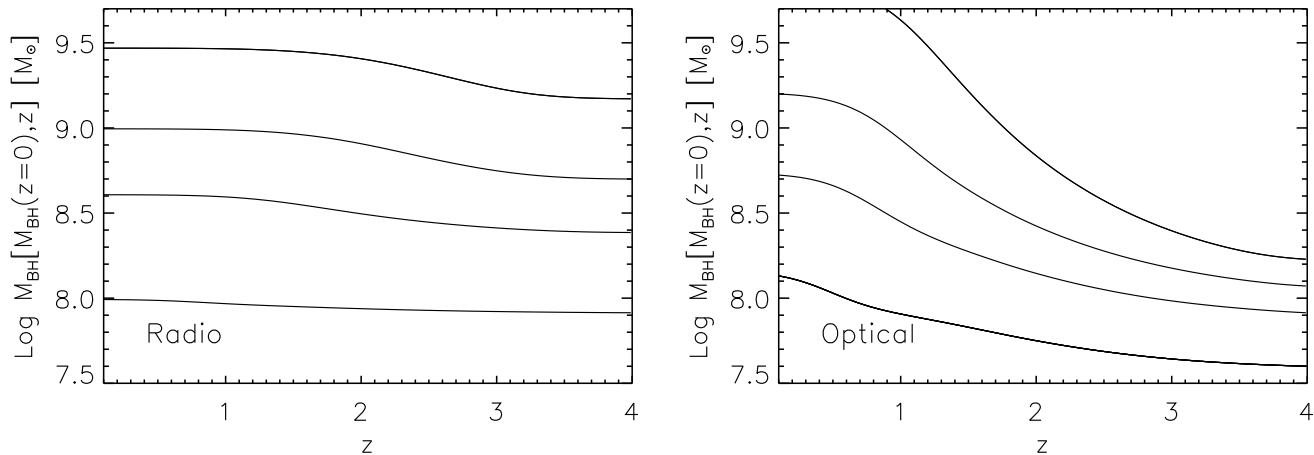}
\caption{Mean accretion growth curves for black
holes of different mass from $z=4$ to $z=0$. \emph{Left panel}: accretion
histories for radio sources alone. \emph{Right panel}: accretion
histories for the optical quasars (see text for further details).
While radio sources only grow by a factor of $\sim 2$, at the most,
optical quasars grow more and at later times.} \label{fig|rhoBHz}
\end{figure*}

Figure~\ref{fig|rhoBHz} shows the mean accretion growth curves for BHs of different mass
from $z=4$ to
$z=0$. The mean BH mass at any time is computed from Equation~(\ref{eq|conteq}) by
integrating the mean accretion rate
\begin{eqnarray}
M_{\rm BH}(M_{\rm BH, i},z)=\int_{z_i}^{z}\langle \dot{M}_{\rm BH}\rangle\frac{dt}{dz}dz\, .
    \label{eq|mdotav}
\end{eqnarray}
The left panel shows the curves of growth for radio sources alone,
while the right panel shows the optical ones. We find that, while optical quasars
can grow up to a factor of 10 along the cosmic evolution from $z=4$ to $z=0$, despite
having a higher Eddington ratio at $z>2$, radio sources
have an average growth not higher than a factor of $\sim 2$. This is due to their
low duty cycle $p_{\rm radio}$, roughly an order of magnitude lower than the optical one. Therefore, although radio sources do accrete at higher \lam\ for a significant amount of time, their overall evolution is still much more
moderate than the optical one. The latter conclusion is strong against possible biases which might affect the exact value of the true underlying Eddington ratio distributions and their evolution with redshift. 

%

Note that here we are not attempting to build a model
for the whole, absolute evolution of all BHs, which would
require a full match to the statistical and clustering
properties of AGNs at all wavelengths
\citep[e.g.,][and references therein]{ShankarCrocce,ShankarReview}.
Here we are just interested to probe the \emph{relative} growth of optical and radio quasars, and to this purpose we
only adopt the optical quasar luminosity function by \citet{Richards05,Richards06},
which is \emph{not} a complete representation of the overall AGN population
\citep[e.g.,][and references therein]{Hop07,SWM}.
Also, the adopted \PL\ distributions are the ones described in \S~\ref{subsec|comparingLambdas}, which may be affected by several biases and uncertainties \citep[e.g.,][]{Shen08bias}.
However, as long as radio and optical quasars are affected by similar
selection effects (see \S~\ref{sec|discu}), the relative comparison is physically meaningful.

\begin{figure*}
\includegraphics[width=17.5truecm]{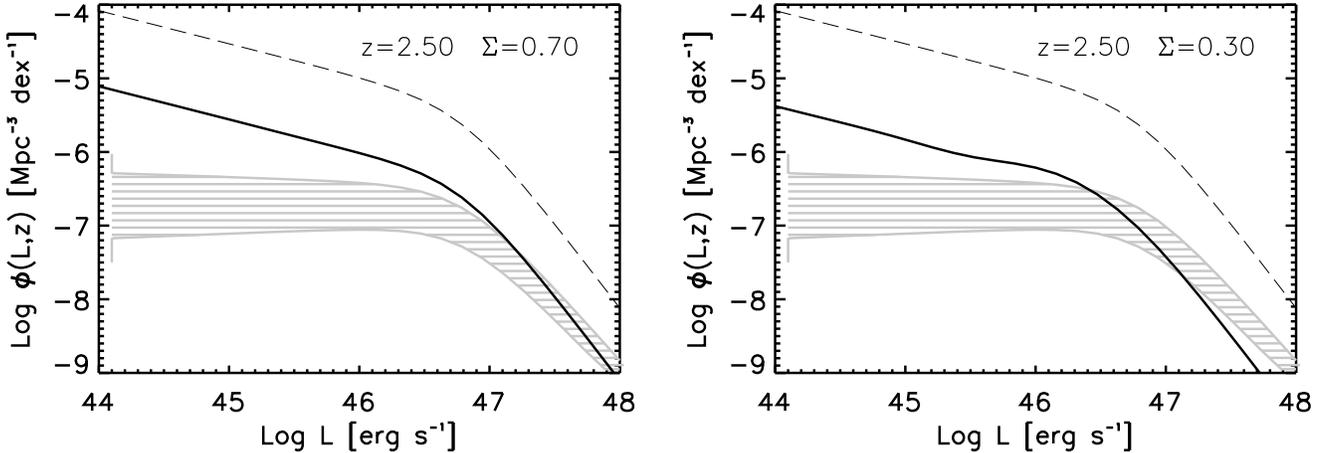}
\caption{\emph{Solid}
lines in both panels are the predicted bolometric luminosity
functions at $z=2.5$ for radio sources alone, obtained by the
convolution of the underlying total black hole mass function with
the assumed \lam-distribution proper for radio sources (see text).
The predicted luminosity functions are compared with gray areas
that mark the optical luminosity function of radio sources alone in
SDSS, expressed in bolometric units, obtained by correcting the
optical quasar luminosity function from Richards et al.\ (2005; shown with \emph{long-dashed} lines) by
the fraction of radio-optical sources measured by Jiang et al.\
(2006) as a function of luminosity and redshift. \emph{Left panel}:
shows the predictions assuming the intrinsic distribution of the
$P(\lambda)$ distribution of radio sources is a Gaussian with
dispersion $\Sigma=0.7$ equal to the optical one. \emph{Right
panel}: same as left panel but assuming $\Sigma=0.3$.
Irrespective
of the exact value for $\Sigma$, lower fractions of radio sources at lower black hole masses
are needed to reproduce the data.}
\label{fig|RadioLF}
\end{figure*}

We now show that we can efficiently test, for radio sources only, if our first assumption of having a
mass-independent underlying \PL\ distribution is a reasonable one.
It is clear that, knowing at each timestep the BH mass function from the continuity equation, and the mapping between (bolometric) luminosity $L$ and BH mass \mbh, provides directly the duty cycle $U(M_{\rm BH},z)$ (see Equation~[\ref{eq|P0general}]).
In other words,
given the observed quasar or radio luminosity function
$\Phi_x(L,z)$, the BH mass function $\Phi_{\rm BH}$, and
the Eddington ratio distribution $p_x$, the duty cycle $U(M_{\rm BH},z)$ can be derived by the equality \citep[e.g.,][]{SteedWeinberg,ShankarReview}
\begin{eqnarray}
\Phi_x(L,z)=\int p_x(\log \lambda,z) U(M_{\rm BH},z)\times \nonumber\\
\Phi_{\rm BH}(M_{\rm BH},z)d\log M_{\rm BH}\, ,
\label{eq|PhiLLambda}
\end{eqnarray}
with $x={\rm opt}$ or $x={\rm radio}$. We apply
Equation~(\ref{eq|PhiLLambda}) to infer the bolometric luminosity
function of radio quasars, given the output duty cycle $U(M_{\rm BH},z)$ and the
underlying assumption that $p_{\rm radio}=0.1\times p_{\rm opt}$, constant for all
BHs of any mass at any time.
Figure~\ref{fig|RadioLF} shows the radio luminosity function predicted from  Equation~(\ref{eq|PhiLLambda}) as
solid lines at a (chosen) redshift of $z=2.5$.
The latter is compared with the gray area, which marks the luminosity function of
radio sources in SDSS at the same redshift, obtained by correcting the quasar
luminosity function from \citet{Richards05} (long-dashed
line in the same Figure) by the luminosity and redshift-dependent radio fraction of
optical sources inferred by \citet{Jiang07}. The left panel shows
the predictions assuming the $p_{\rm radio}(\lambda,z)$ distribution of radio sources in
Figure~\ref{fig|PLambda} to be constant with BH mass. It can be seen
that the predictions overproduce the actual observed radio
luminosity function at the faint end, which implies that the input
\PL\ is \emph{not} correct. The right panel of
Figure~\ref{fig|RadioLF} shows the result of a similar exercise in
which we instead insert a $p_{\rm radio}(\lambda,z)$ distribution in the
continuity equation with a much narrower intrinsic scatter of
$\Sigma=0.3$. The $z=2.5$ predictions for the latter model imply now
less radio sources at a given luminosity as the overall probability
$p_{\rm radio}(\lambda,z)$ is narrower, an effect which decreases the
probability for BHs to be active as radio sources. However, it can
be seen that even the latter model provides a poor match to the
data. We conclude that, irrespective of the exact value for the
broadness of the $p_{\rm radio}(\lambda,z)$ distribution, the only way to
reproduce the observed fraction of radio sources in the faint end of the quasar
luminosity function, is to assume $p_{\rm radio}(\lambda,z)=k(M_{\rm BH}) p_{\rm opt}(\lambda,z)$, with $k(M_{\rm BH})$ being significantly lower than 10\% at lower BH masses. This would produce an increasingly lower fraction of
BHs as active radio sources at lower masses, and a lower number of
radio sources at fainter bolometric luminosities. These findings are in line
with the results derived in local galaxies by \citet{Best05}, who
claim a similar, or even steeper, decline of the AGN fraction with
decreasing BH mass.



\section{DISCUSSION}
\label{sec|discu}

\subsection{LOOKING FOR BIASES}
\label{subsec|lookingforbiases}

The SDSS is flux-limited, and therefore one might argue that the
possibly heavy loss of faint sources might bias our results on a
different accretion history between optical and radio sources.
However, Figures~\ref{fig|LambdaDistributionsLowHighMass} and
\ref{fig|MbhDistributionsLowHighLambda} show that even if we
restrict our analysis to the subsample of BHs with mass \mbh$\gtrsim
10^9\,$\msun, which always tend to shine
above the SDSS flux limit (see Figure~\ref{fig|initialSummary}),
we find very similar Eddington ratio and BH mass
distributions as in the total sample. In principle, massive BHs accreting
at very low Eddington ratios should be missed in SDSS, thus possibly biasing the above result. However, we also note that the $z>2$ median BH mass is \mbh$\sim2\times 10^9$\msun\ radiating at \lam$\sim 0.4$, and SDSS would be able to detect them radiating down to \lam$=0.1$, i.e. $L_{\rm BOL}\sim 2\times 10^{46}\,$\ergs\ (cf. Figure~\ref{fig|initialSummary}).

Nevertheless, the results discussed in this paper may be
significantly affected by other biases and measurement errors. For
example, \citet{Shen09} have recently discussed several selection
biases which may underestimate the mean and broadness of the
Eddington ratio distributions in flux limited samples. Also, the
\CIV\ lines may be affected by winds and therefore the
masses of high redshift quasars might be overestimated with respect to the \Mgg\ based ones. \citet{Shen09} discuss that \CIV\ masses are correlated with
the \Mgg\ ones, although with a slight offset and much larger
scatter. Analogously, the analysis of fluxes in the DR6 SDSS sample
has shown that all fluxes may be systematically underestimated in
the DR3 sample. Last but not least, even if reverberation mapping virial
relations are (strongly) biased \citep[e.g.,][but see also Netzer 2008]{Marconi08}, this does not adversely affect our analysis or results.
In fact, although all the above biases
may induce strong uncertainties in the absolute
measurements of BH masses or AGN bolometric luminosities, there is
no obvious reason why they should affect radio and optical quasars
in a different way. Therefore the relative, systematic offsets
between radio and optical sources discussed in \S~\ref{sec|results},
should be reliable. Moreover, none of the effects listed above would
be capable of inducing the \emph{redshift-dependent}
differences observed in the accretion histories of the two quasar
populations.

A more subtle bias may arise from a different underlying mass
distribution for optical and radio active BHs. In flux-limited
samples, lower mass BHs with steeper number distribution are
scattered into higher mass bins more efficiently than those in
flatter number distribution, thus biasing the Eddington ratio
distributions.
For narrow Eddington ratio distributions
and bright luminosities, the active BH mass function has
a similar shape to the AGN luminosity function (see \citealt{SWM}), as also
empirically found via direct calibration by \citet{Veste08}.
Therefore, a direct comparison of the optical and radio quasar luminosity
function bright-end slopes, is similar to comparing the
mass distributions of active BHs.
To this purpose, we have multiplied the
\citet{Richards05} luminosity function for the usual,
completeness-corrected radio fraction of \citet{Jiang07} to yield
an optical luminosity function for radio sources alone. We find
that, at all redshifts of interest here, the bright end slope for
the radio-loud quasar luminosity function
is always shallower than the optical one (note that the
specific value of the bright end slope of the quasar luminosity
function is irrelevant for this test). This in turn would imply
that BH masses for optical sources could be smaller and their intrinsic
Eddington ratios higher. While this effect might play some role in the
behaviour seen at $z<2$, it would certainly not be able to explain
the opposite behaviour seen at higher redshifts, where the effects due to the flux
limits should be, if anything, even stronger.

A more physical bias may derive from the fact that our radio
sampling is restricted to optically selected sources, and many more
sources are found in high frequency radio surveys which may not have
counterparts in SDSS \citep[e.g.,][and references therein]{Ciras03,Dezotti05,Massardi08,Dezotti09}.
Nevertheless, although radio activity in AGNs is still not well
understood and may pass through different stages
\citep[e.g.,][]{Croton06,Blundell07,Heinz07,Cav08,GhiselliniTavecchio,MerloniHeinz08,Shankar08Cav},
it is clear that at least within luminous, optically selected
sources, radio-loudness is not a simple function of BH mass or
Eddington ratio, and that radio sources are not a mere random
subsample of the optical ones.


\subsection{HINTS FROM CLUSTERING}
\label{subsec|clustering}

We find that, since $z=4$, the accretion histories of optical and
radio source are significantly different, in a non trivial and
redshift-dependent way.
This suggests that powerful radio
and optical sources may be intrinsically different. Independent
empirical studies on the clustering properties of optical and radio
sources support our results. \citet{Negrello06} find that the
observed two-point angular correlation function of milliJansky radio
sources exhibits the puzzling feature of a power-law behaviour up to
very large ($\sim 10^{\circ}$) angular scales which cannot be
accounted for in the standard hierarchical clustering scenario for
any realistic redshift distribution of such sources. The radio
sources responsible for the large-scale clustering signal are
increasingly less clustered with increasing look-back time, up to at
least $z\sim$1, at variance with what found for optically selected
quasars \citep[e.g.,][]{Croom05,Porciani06}. The data are accurately
accounted for in terms of a bias function which decreases with
increasing redshift, mirroring the evolution with cosmic time of the
characteristic halo mass entering the non-linear regime. More
recently, \citet{Shen09} found that radio-loud quasars are more
strongly clustered than radio-quiet quasars of similar mass. This
implies that radio-loud quasars live in more massive dark matter
haloes and denser environments than radio-quiet quasars, consistent
with local $z < 0.3$ observations for radio-loud type 2 AGNs
\citet{Mandelbaum08} and radio galaxies \citep{Lin08,Wake08}. Also
the hosts of optical and radio quasars seem to have somewhat
different structural properties \citep[e.g.,][]{Wolf08}.

\subsection{IMPLICATIONS}
\label{subsec|implications}

The high BH mass, high redshift radio sources observed in our sample might play a significant role in preheating the cores of groups and clusters \citep[e.g.,][and references therein]{Bower08,Cav08,MerloniHeinz08}. Completing their growth
already at $z\gtrsim 2$, these massive BHs can in fact induce radiative and
kinetic energy in their surroundings already at very early epochs, thus significantly contributing to increasing
the entropy in their surroundings \citep[e.g.,][]{Cav08}.
Moreover, if the radio AGN phenomenon is preferentially confined within the subsample
of the optical quasars which tend to live
in overdense environments (e.g., groups and clusters),
the AGN radio feedback will prevent the ionization of lower density regions of the universe.
This possibility might be reconciled with independent studies that find that any injection of non-gravitational energy in the diffuse baryons should avoid low-density regions at high redshift to be consistent with the void statistics of the $z\sim 2$ observed Lyman $\alpha$-forest \citep[e.g.,][]{Borgani08}.

It has been often discussed in the Literature that a high BH mass
may be a necessary although not sufficient, condition for AGN
radio-loudness \citep[e.g.,][]{Laor00,Ho02,Best05,Gopal08}. A more recent study
by \citet{Rafter09} suggests that, although there is indeed a tendency
for the more massive BHs to have a higher probability of being active radio sources, no clear demarcation is apparent below a BH mass of $\sim 2\times 10^8$\msun. 
The results presented in
Figure~\ref{fig|SummaryMbhvsRedshiftHighLowLambda} show that
radio-loud quasars do actually cover a wide range of BH masses, and that
their masses are offset with respect to those of optical quasars.
More specifically, although radio-loud quasars are, on average, always characterized by higher BH masses, their offset with respect to optical quasars
steadily decreases with increasing redshift, ending up at $z>2$ having
comparable, or even
\emph{lower}, BH masses than
optical quasars.
Our data therefore do not point to any clear trend between radio-loudness and specific BH mass. Instead, our analysis seems to suggest that radio and optical quasars
have different accretion
histories. While at most times radio-loud quasars are preferentially
associated with more massive BHs than radio-quiet quasars, there is
no clear dividing line in BH mass.
However, as we show above from detailed evolutionary models, radio-loud quasars have not grown
their mass by a significant amount since $z\sim 4$, mainly due to their low
duty cycles. Thus, the massive
radio-loud quasars observed at late times must have grown their mass at earlier epochs than those probed here (see also Overzier et al. 2009 for a similar conclusion
on the rapid growth of $z>4$ radio galaxies).

Several groups have also put forward the possibility of a similarity
between X-ray/radio galactic binaries and AGNs
\citep[e.g.,][]{Meier01,Gallo03,Maccarone03,MerloniFP,Falcke04,Fender04,Jester05}.
In particular, it has been shown that there might be a common
scaling relation between X-ray luminosity, radio luminosity, and BH
mass in X-ray binaries and AGNs \citep{MerloniFP}. Different
observational states have been observed for X-ray binaries. In
brief, X-ray sources at very low Eddington ratios $\lesssim
10^{-2}$, are observed to be inefficient optical emitters, but
efficient radio-jet emitters, and are therefore defined to be in a
``low (luminosity)/hard (spectrum)'' (or power-law) state. At higher Eddington
ratios X-ray binaries are observed in a thermal, radiatively
efficient, disk-dominated phase. When X-ray binaries enter this
thermal ``soft'' state the steady jet is quenched
\citep[e.g.,][]{Gallo03}. A second transition occurs at Eddington
ratios higher than 30\% when X-ray binaries enter a ``very high
state'' with a steep power law spectrum and intermittent radio-jet
activity. Given the similarities in accretion physics, it is
tempting to associate similar states to AGNs. However, the results
presented in Figure~\ref{fig|SummaryLambdavsBHmassLambdavsRedshift}
may pose serious problems to the connection between X-ray binaries
and AGNs. In fact, radio sources in our sample encompass a
significantly large range of Eddington ratios with no clear evidence
of transition thresholds. Interestingly, \citet{Maoz07} also finds
that a group of radio-loud LINERS, which are thought to be
radiatively inefficient sources with no ``big blue bump'', shows
instead a spectral energy distribution very similar to that of
Seyfert galaxies, which require thin accretion disks.
Also, several empirical works suggested \citep[e.g.,][]{Churazov01}, also in analogy to X-ray binaries and microquasars \citep[e.g.,][]{Reynolds97,Nipoti05}, that
radio activity might be a brief and ``intermittent'' phase, tuned in a way to yield the low fraction of radio sources observed within optical samples.
Although our study does not allow any definite constraint on such intermittency, it does however suggest that even if radio emission is intermittent, the cycles
are not distributed randomly in time and mass.

\citet{Sikora07} collected a significant sample of
radio-loud and radio-quiet quasars, spanning a large range of
Eddington ratios. They find that radio-loud sources define an upper
sequence in the radio-loudness versus Eddington ratio plane, suggesting that
there is no clear correlation between radio-loudness and BH mass or \lam.
Overall, we also find no significant connection between radio activity
and BH mass and/or accretion rate, a result which may indicate, in
agreement with \citet{Sikora07}, that other BH properties, such as
the spin, may be responsible for the radio activity in some AGNs. In
their analytic model, \citet{WilsonColbert} were able to reproduce
the radio luminosity function by assuming that radio-loud quasars
are a different, non-random subsample of optical quasars
characterized by a higher spin. More recently, \citet{Lagos09}
adopting a full model for galaxy and BH evolution, found that the
final BH spin distribution depends almost exclusively on the BH
accretion history, with the main mechanisms of BH spin-up being gas
cooling processes and disc instabilities. They found that the more
massive BHs, which are hosted by massive elliptical galaxies, have higher
spin values than less-massive BHs, hosted by spiral galaxies.
Similar results were also claimed by \citet{Volonteri07}, who found
that the observed radio loudness bimodality is directly related to
the BH spin distribution in galaxies. In their model, BHs in giant
elliptical galaxies are grown by merger-driven
accretion and end up having, on average, much larger spins than BHs
in spiral, disk galaxies.


\section{CONCLUSIONS}
\label{sec|conclu}

In this paper we have cross-correlated the SDSS DR3 sample with
FIRST and with the \citet{Veste08} BH mass sample. We found significant
statistical evidence for the radio sources to have a higher \lam\
at $z>2$ with respect to optical quasars. The
situation reverses at $z<1$, where radio sources have lower
Eddington ratios. At $z > 2$ radio quasars tend to be less massive
than optical quasars; however, as redshift decreases radio quasars happen to be
in increasingly more massive BHs with respect to optical quasars. We have checked that all
these results cannot be a result of any evident bias.
For example,
restricting to the subsample of active BHs with mass $\gtrsim
10^9\,$\msun, which is not affected by flux-limited effects, yields
essentially equal results. Also, we have discussed that any other
bias, such as systematic uncertainties in SDSS fluxes, in BH mass
measurements, or different slopes in the intrinsic active mass
function of radio and optical quasars, are not able to induce the
\emph{redshift-dependent} systematic differences we observe
between radio and optical quasars
in the
SDSS data. Our results suggest that optical and optically-selected
radio sources have different accretion histories since very early
epochs, and may be hosted by different dark matter haloes, as also
suggested by some clustering measurements. We find no clear
correlation between radio activity and BH mass and/or accretion
rate in our data, which may hint towards another BH property as source of radio
activity, such as the BH spin. We perform detailed modelling of the
accretion histories of optical and radio sources in terms of a
continuity equation and broad input Eddington ratio distributions.
We find that while optical sources may grow up to an order of
magnitude, radio sources had a much more contained growth since $z\sim 2-4$.
The same
modelling allows us to conclude that the probability for lower mass
BHs to be radio loud must be lower than for higher mass BHs at
all epochs, to reproduce the low fraction of radio sources
at faint optical luminosities as observed in SDSS.

\section*{Acknowledgments}
FS acknowledges the Alexander von Humboldt Foundation and NASA Grant NNG05GH77G for support. MV acknowledges financial support through grants HST-AR-10691, HST-GO-10417, and HST-GO-10833 from NASA through the Space Telescope Science Institute, which is operated by the Association of Universities for Research in Astronomy, Inc., under NASA contract NAS5-26555.
We thank Guinevere Kauffmann, Philip Best, Marek Sikora, Marta Volonteri, Elena Gallo, and Eyal Neistein, for interesting discussions.

\begin{table*}
\caption{Eddington Ratio Distributions of Optical and Radio Quasars\label{tab|eddratio}}
\begin{center}
\begin{tabular}{lrrrrrcrrrcrr}
\hline
\hline
&
\multicolumn{2}{c}{Ranges}&
&
\multicolumn{3}{c}{Optical Quasars (O)}&
&
\multicolumn{3}{c}{Radio Quasars (R)}&
&
\\
\cline{2-3}
\cline{5-7}
\cline{9-11}
\multicolumn{1}{l}{Row}&
\multicolumn{1}{c}{z} &
\multicolumn{1}{c}{log \mbh} &
&
\multicolumn{1}{c}{$N$} &
\multicolumn{1}{c}{$\mu(\log \lambda)$} &
\multicolumn{1}{c}{$\sigma(\log \lambda)$} &
&
\multicolumn{1}{c}{$N$} &
\multicolumn{1}{c}{$\mu(\log \lambda)$} &
\multicolumn{1}{c}{$\sigma(\log \lambda)$} &
\multicolumn{1}{c}{$\mu_{O}-\mu_{\rm R}$} &
\multicolumn{1}{c}{$P_{T}$}
\\
\hline
1  & [0.09,1.00) & [6.59,\phn9.00) & & 2705 & $-0.815\pm{0.008}$ & $0.40$ & &  169 & $-0.905\pm{0.036}$ & $0.47$ & $ 0.089\pm{0.037}$ & 2.1E-02\\
2  & [1.00,1.50) & [6.59,\phn9.00) & & 1899 & $-0.502\pm{0.006}$ & $0.28$ & &   97 & $-0.419\pm{0.031}$ & $0.31$ & $-0.083\pm{0.032}$ & 9.0E-03\\
3  & [1.50,2.00) & [6.59,\phn9.00) & & 1362 & $-0.306\pm{0.007}$ & $0.26$ & &   62 & $-0.212\pm{0.036}$ & $0.29$ & $-0.094\pm{0.037}$ & 1.6E-02\\
4  & [2.00,4.75) & [6.59,\phn9.00) & &  596 & $-0.086\pm{0.014}$ & $0.33$ & &   59 & $ 0.053\pm{0.038}$ & $0.29$ & $-0.139\pm{0.040}$ & 8.8E-04\\ 
\\
5  & [0.09,1.00) & [9.00,10.30)    & &  338 & $-1.253\pm{0.016}$ & $0.30$ & &   80 & $-1.321\pm{0.037}$ & $0.33$ & $ 0.068\pm{0.040}$ & 7.5E-02\\
6  & [1.00,1.50) & [9.00,10.30)    & & 1250 & $-0.923\pm{0.007}$ & $0.25$ & &  136 & $-0.950\pm{0.029}$ & $0.34$ & $ 0.027\pm{0.030}$ & 3.9E-01\\
7  & [1.50,2.00) & [9.00,10.30)    & & 1967 & $-0.632\pm{0.005}$ & $0.24$ & &  108 & $-0.566\pm{0.031}$ & $0.32$ & $-0.066\pm{0.031}$ & 2.6E-02\\
8  & [2.00,4.75) & [9.00,10.30)    & & 2255 & $-0.547\pm{0.005}$ & $0.25$ & &  117 & $-0.361\pm{0.028}$ & $0.30$ & $-0.186\pm{0.029}$ & 4.3E-10\\
\\
9  & [0.09,1.50) & [6.59,\phn8.00) & &  547 & $-0.558\pm{0.014}$ & $0.32$ & &   25 & $-0.632\pm{0.061}$ & $0.31$ & $ 0.074\pm{0.063}$ & 3.5E-01\\
10 & [0.09,1.50) & [8.00,\phn8.50) & & 1433 & $-0.615\pm{0.011}$ & $0.43$ & &   80 & $-0.617\pm{0.055}$ & $0.49$ & $ 0.002\pm{0.056}$ & 9.9E-01\\
11 & [0.09,1.50) & [8.50,\phn9.00) & & 2624 & $-0.733\pm{0.007}$ & $0.35$ & &  161 & $-0.773\pm{0.038}$ & $0.48$ & $ 0.040\pm{0.039}$ & 2.4E-01\\
12 & [0.09,1.50) & [9.00,\phn9.50) & & 1533 & $-0.981\pm{0.007}$ & $0.29$ & &  191 & $-1.082\pm{0.026}$ & $0.36$ & $ 0.101\pm{0.027}$ & 6.4E-04\\
13 & [0.09,1.50) & [9.50,10.30)    & &   55 & $-1.123\pm{0.043}$ & $0.32$ & &   25 & $-1.114\pm{0.096}$ & $0.48$ & $-0.008\pm{0.105}$ & 8.9E-01\\
\\
14 & [1.50,4.75) & [6.59,\phn8.00) & &    4 & $ 0.917\pm{0.095}$ & $0.19$ & &    1 & $ 0.755\pm{\rm \cdots \ \ \  }$ & $\cdots$ & $\cdots$ & $\cdots$\\
15 & [1.50,4.75) & [8.00,\phn8.50) & &  180 & $ 0.208\pm{0.023}$ & $0.31$ & &   13 & $ 0.139\pm{0.083}$ & $0.30$ & $ 0.069\pm{0.086}$ & 3.8E-01\\
16 & [1.50,4.75) & [8.50,\phn9.00) & & 1774 & $-0.284\pm{0.006}$ & $0.27$ & &  107 & $-0.114\pm{0.029}$ & $0.30$ & $-0.170\pm{0.030}$ & 9.4E-08\\
17 & [1.50,4.75) & [9.00,\phn9.50) & & 3389 & $-0.570\pm{0.004}$ & $0.25$ & &  182 & $-0.440\pm{0.024}$ & $0.32$ & $-0.130\pm{0.024}$ & 7.4E-08\\
18 & [1.50,4.75) & [9.50,10.30)    & &  833 & $-0.657\pm{0.009}$ & $0.25$ & &   43 & $-0.536\pm{0.055}$ & $0.36$ & $-0.121\pm{0.056}$ & 2.1E-02\\
\\
19 & [0.09,1.00) & [6.59,\phn9.00) & & 2705 & $-0.815\pm{0.008}$ & $0.40$ & &  169 & $-0.905\pm{0.036}$ & $0.47$ & $ 0.089\pm{0.037}$ & 2.1E-02\\
20 & [1.00,1.50) & [6.59,\phn9.00) & & 1899 & $-0.502\pm{0.006}$ & $0.28$ & &   97 & $-0.419\pm{0.031}$ & $0.31$ & $-0.083\pm{0.032}$ & 9.0E-03\\
21 & [1.50,2.00) & [6.59,\phn9.00) & & 1362 & $-0.306\pm{0.007}$ & $0.26$ & &   62 & $-0.212\pm{0.036}$ & $0.29$ & $-0.094\pm{0.037}$ & 1.6E-02\\
22 & [2.00,4.75) & [6.59,\phn9.00) & &  596 & $-0.086\pm{0.014}$ & $0.33$ & &   59 & $ 0.053\pm{0.038}$ & $0.29$ & $-0.139\pm{0.040}$ & 8.8E-04\\
\\
23 & [0.09,1.00) & [9.00,10.30)    & &  338 & $-1.253\pm{0.016}$ & $0.30$ & &   80 & $-1.321\pm{0.037}$ & $0.33$ & $ 0.068\pm{0.040}$ & 7.5E-02\\
24 & [1.00,1.50) & [9.00,10.30)    & & 1250 & $-0.923\pm{0.007}$ & $0.25$ & &  136 & $-0.950\pm{0.029}$ & $0.34$ & $ 0.027\pm{0.030}$ & 3.9E-01\\
25 & [1.50,2.00) & [9.00,10.30)    & & 1967 & $-0.632\pm{0.005}$ & $0.24$ & &  108 & $-0.566\pm{0.031}$ & $0.32$ & $-0.066\pm{0.031}$ & 2.6E-02\\
26 & [2.00,4.75) & [9.00,10.30)    & & 2255 & $-0.547\pm{0.005}$ & $0.25$ & &  117 & $-0.361\pm{0.028}$ & $0.30$ & $-0.186\pm{0.029}$ & 4.3E-10\\
\hline
\end{tabular}
\end{center}
\raggedright Notes: For each specified range, this table lists the number of optical and radio quasars
$N$, and the (biweight) mean $\mu$ and (biweight) standard deviation $\sigma$ for
distributions of the logarithm of the Eddington ratio. Both the difference of the means $\mu_{O}-\mu_{\rm
R}$ and the statistical chance the optical and radio quasars have the same mean
as determined by the Student's T statistic $P_T$ are listed.
\end{table*}

\begin{table*}
\caption{Black Hole Mass Distributions of Optical and Radio Quasars\label{tab|bhmass}}
\begin{center}
\begin{tabular}{lrrrrrcrrrcrr}
\hline
\hline
&
\multicolumn{2}{c}{Ranges}&
&
\multicolumn{3}{c}{Optical Quasars (O)}&
&
\multicolumn{3}{c}{Radio Quasars (R)}&
&
\\
\cline{2-3}
\cline{5-7}
\cline{9-11}
\multicolumn{1}{l}{Row}&
\multicolumn{1}{c}{z} &
\multicolumn{1}{c}{log \lam} &
&
\multicolumn{1}{c}{$N$} &
\multicolumn{1}{c}{$\mu(\log M_{\rm BH})$} &
\multicolumn{1}{c}{$\sigma(\log M_{\rm BH})$} &
&
\multicolumn{1}{c}{$N$} &
\multicolumn{1}{c}{$\mu(\log M_{\rm BH})$} &
\multicolumn{1}{c}{$\sigma(\log M_{\rm BH})$} &
\multicolumn{1}{c}{$\mu_{O}-\mu_{\rm R}$} &
\multicolumn{1}{c}{$P_{T}$}
\\
\hline
1     & [0.09,1.00) & [-2.7,-0.6) & & 2197 & $ 8.577\pm{0.009}$ & $0.43$ & &  210 & $ 8.776\pm{0.035}$ & $0.51$ & $-0.199\pm{0.036}$ & 1.4E-07\\
2     & [1.00,1.50) & [-2.7,-0.6) & & 1829 & $ 9.058\pm{0.005}$ & $0.23$ & &  149 & $ 9.194\pm{0.021}$ & $0.26$ & $-0.137\pm{0.022}$ & 1.5E-08\\
3     & [1.50,2.00) & [-2.7,-0.6) & & 1249 & $ 9.212\pm{0.005}$ & $0.19$ & &   53 & $ 9.276\pm{0.026}$ & $0.19$ & $-0.064\pm{0.026}$ & 2.2E-02\\
4     & [2.00,4.75) & [-2.7,-0.6) & &  982 & $ 9.459\pm{0.007}$ & $0.23$ & &   24 & $ 9.523\pm{0.054}$ & $0.27$ & $-0.064\pm{0.055}$ & 1.9E-01\\
\\
5     & [0.09,1.00) & [-0.6, 1.4) & &  846 & $ 8.143\pm{0.014}$ & $0.40$ & &   39 & $ 8.325\pm{0.080}$ & $0.50$ & $-0.182\pm{0.081}$ & 3.2E-02\\
6     & [1.00,1.50) & [-0.6, 1.4) & & 1320 & $ 8.649\pm{0.008}$ & $0.29$ & &   84 & $ 8.765\pm{0.038}$ & $0.35$ & $-0.116\pm{0.039}$ & 1.4E-03\\
7     & [1.50,2.00) & [-0.6, 1.4) & & 2080 & $ 8.945\pm{0.006}$ & $0.26$ & &  117 & $ 8.994\pm{0.027}$ & $0.29$ & $-0.049\pm{0.027}$ & 8.5E-02\\
8     & [2.00,4.75) & [-0.6, 1.4) & & 1869 & $ 9.163\pm{0.008}$ & $0.36$ & &  152 & $ 9.100\pm{0.030}$ & $0.37$ & $ 0.062\pm{0.031}$ & 3.4E-02\\
\\
9$^a$ & [0.09,1.00) & [-2.7,-0.6) & &  869 & $ 8.958\pm{0.007}$ & $0.20$ & &  116 & $ 9.122\pm{0.024}$ & $0.26$ & $-0.164\pm{0.025}$ & 2.8E-09\\
10$^a$& [1.00,1.50) & [-2.7,-0.6) & & 1723 & $ 9.081\pm{0.005}$ & $0.21$ & &  142 & $ 9.216\pm{0.019}$ & $0.23$ & $-0.135\pm{0.020}$ & 1.2E-10\\
11$^a$& [1.50,2.00) & [-2.7,-0.6) & & 1249 & $ 9.212\pm{0.005}$ & $0.19$ & &   53 & $ 9.276\pm{0.026}$ & $0.19$ & $-0.064\pm{0.026}$ & 2.2E-02\\
12$^a$& [2.00,4.75) & [-2.7,-0.6) & &  976 & $ 9.461\pm{0.007}$ & $0.23$ & &   24 & $ 9.523\pm{0.054}$ & $0.27$ & $-0.062\pm{0.055}$ & 2.2E-01\\
\\
13$^a$& [0.09,1.00) & [-0.6, 1.4) & &   53 & $ 8.836\pm{0.017}$ & $0.12$ & &   10 & $ 8.886\pm{0.038}$ & $0.12$ & $-0.051\pm{0.041}$ & 2.8E-01\\
14$^a$& [1.00,1.50) & [-0.6, 1.4) & &  559 & $ 8.903\pm{0.007}$ & $0.16$ & &   49 & $ 8.989\pm{0.036}$ & $0.25$ & $-0.086\pm{0.037}$ & 2.0E-02\\
15$^a$& [1.50,2.00) & [-0.6, 1.4) & & 1726 & $ 9.018\pm{0.005}$ & $0.20$ & &  102 & $ 9.056\pm{0.022}$ & $0.23$ & $-0.038\pm{0.023}$ & 8.9E-02\\
16$^a$& [2.00,4.75) & [-0.6, 1.4) & & 1667 & $ 9.229\pm{0.007}$ & $0.29$ & &  131 & $ 9.183\pm{0.026}$ & $0.29$ & $ 0.046\pm{0.027}$ & 7.0E-02\\

\hline
\end{tabular}
\end{center}
\raggedright Notes: For each specified range, this table lists the number of optical and radio quasars
$N$, and the (biweight) mean $\mu$ and (biweight) standard deviation $\sigma$ for
distributions of the  logarithm of the black hole mass. Both the difference of the means $\mu_{O}-\mu_{\rm
R}$ and the statistical chance the optical and radio quasars have the same mean
as determined by the Student's T statistic $P_T$ are listed.\\
$^a$: log \mbh$>8.7$ was also required.
\end{table*}

\bibliographystyle{mn2e}
\bibliography{../../RefMajor}

\label{lastpage}

\end{document}